\documentclass[journal,twoside]{IEEEtran}

\usepackage{cite}

\ifCLASSINFOpdf
  \usepackage[pdftex]{graphicx}
  \DeclareGraphicsExtensions{.pdf,.jpeg,.png}
\else
\fi
\usepackage{amsmath}
\usepackage{array}

\usepackage[caption=false,font=footnotesize]{subfig}
\begin{document}
%
\title{A 1V 5-bits Low Power  Level Crossing ADC with OFF state in idle time for bio-medical applications in $0.18 \mu m$ CMOS}
%
%
%

\author{Lucas~Moura~Santana,~\IEEEmembership{Student,~IEEE,}
        Duarte~Lopes~de~Oliveira,~\IEEEmembership{Member,~IEEE,}
        and~Lester~de~Abreu~Faria,~\IEEEmembership{Member,~IEEE}
\thanks{Footnote}
\thanks{L. M. Santana, D. L. de Oliveira and L. A. Faria are within the Technological Institute of Aeronautics.}%
\thanks{Manuscript written on April 6, 2019;}}

%
%

\markboth{IEEE Transaction on Circuits and Systems I: Regular Papers.}%
{Santana \MakeLowercase{\textit{et al.}}: A 1V 5-bits Low Power  Level Crossing ADC with OFF state in idle time for bio-medical applications in $0.18 \mu m$ CMOS}
%



\maketitle

\begin{abstract}
The ubiquitous use of sensing and signal processing is increasing exponentially with the advance of the Internet of Everything (IoE). In this context, the design of every time more power efficient sensor nodes is a must. Within these nodes, one of the most power-hungry components are the analog-to-digital converters (ADC). These components are used everywhere to translate real-world analog signals into computer intelligible digital signals. One of the promising architecture for the sensing of physiological signals is the level crossing ADC due to the sparse characteristics of those signals. One of the challenges to improve the power efficiency of this type of ADC lies in the use of continuous comparators to keep track of the input signal within the voltage references. The aim of this work is to investigate the impact of using continuous comparator which can be turned off without incurring error to the conversion of the level crossing ADC. New boundaries will be set for the correct behavior of the level crossing ADC together with the conditions for power saving with the proposed architecture. A 1V 5-bits level crossing ADC was implemented using the TSMC $0.18 \mu m$ process and fabricated for laboratory measurements. The ADC consumes $12.2\mu W$ during tracking state and with the proposed technique, the reduction of the average power can go from 4.2\% to 45.5\% depending on the activity and the type of the input signal. 
\end{abstract}


%
\IEEEpeerreviewmaketitle

\section{Introduction}
\IEEEPARstart{T}{he} data compression method using level crossing shows some interesting features since it was proposed in the construction of medium resolution (4 to 10 bits) and low bandwidth Analog-to-Digital converters (ADC) \cite{mark1981, Sayiner1992}. From the performance perspective, the level crossing ADC has the advantage of using the least amount of data to provide the output, being equal to two bits, one for the direction of the level crossing and another for the occurrence of a level crossing. Those two bits combined with an initial condition are able to provide any number of bits at the output, given the proper behavior of the ADC when tracking the input signal. Also the asynchronous behavior of the sampling, given it is input dependent, indicates the level crossing ADC as a good solution for the conversion of sparse signals, which are signals with high spectrum spike contents followed by long idle periods \cite{Sayiner1992}.

Most of the recent work on level crossing ADCs are being made to lower the power consumption of such converter while keeping the medium resolution to use it as a feasible solution for bio-medical applications (\cite{Zhang2014, Li2013, Marisa2017}). In this work, a technique to further reduce the power consumption under a known input signal conditions is evaluated. This technique consists of turning off the comparators for signal tracking while those are not being used. A set of conditions for the application of such technique are given for the proper functioning of the ADC under a well-behaved input signal. A prototype 5-bits level crossing ADC is fabricated in $0.18\mu m$ to prove the operation of the conversion. The final ADC consumes a tracking power of $12.2\mu W$ and such power can be reduced by a 45.5\% for a pure sine wave input and between 4.2\% to 32.7\% for a scaled ECG signal.

This work is then divided as follow for further explanation. Section II is presenting the level crossing architecture implemented. Section III is describing the design of the key components such as comparators and the finite state machine, together with the implementation of the power reduction technique. Section IV is about the measurements of the fabricated ADC and the analysis of the power reduction technique. Finally, Section V concludes the paper and presents further steps to improve the ADC power reduction.

\section{Architecture Review}
To understand the behavior of the general level crossing ADC one can approximate this type of conversion to the one of two other types of ADC: the flash and the Successive Approximation Register (SAR). The level crossing is similar to the flash since it continuously generates voltage references that feed comparators to decide the actual conversion output of the ADC. The similarity with the SAR is in the aspect of the sampling event triggering a binary search, that is, every time a conversion is being made the system will take action to update its values if the level crossing is going up or if it is going down.

\begin{figure*}[!t]
    \centering
    \subfloat[Floating window generic implementation]{\includegraphics[width=0.5\linewidth]{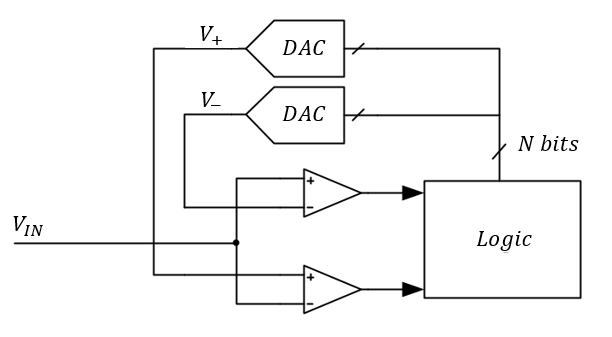}%
    	\label{float_architecture}}
    \hfil
    \subfloat[Floating window tracking with the moving voltage references]{\includegraphics[width=0.5\linewidth]{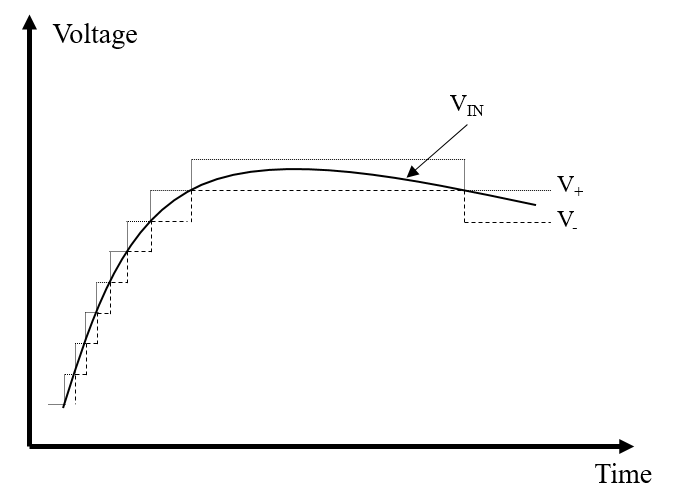}%
    	\label{floating_window}}
    \caption{Floating window architecture and tracking of the input signal.}
    \label{floating_set}
\end{figure*}

The implementation of level crossing ADCs is done mainly by two different architectures: the floating window and the fixed window \cite{Li2015}. A window may be seen as the space between two reference voltages where the input signal is located at a given time. The event of the signal going out of the window triggers a level crossing, thus the update of the system to accommodate the window to the input signal again. The difference in voltage between the references is always set to be one quantization step of the ADC or one LSB. For its simplicity and robustness, the floating window is used in this work and it is further explained. Figure \ref{floating_set} shows the simplest way to implement the floating window level crossing architecture and it depicts the moving window tracking the input signal.

As for choosing the implementation of the components presented in the architecture, there lie the many variations of the level crossing ADCs. For the DACs, the work in \cite{Zhang2014} uses a resistive ladder for its flexibility in generating the moving window; \cite{Li2013} uses an innovative 1-bit CDAC which is able to perform with great efficiency in the fixed window scheme; \cite{Marisa2017} implements a set of analog memories to add the capability to easily change the resolution of the ADC accordingly. For the logic that keeps track of the crossing event, there are also many ways to control RTL components such as registers and ALUs. However, when reaching the comparator design phase, one could only choose among the topologies within the continuous comparators. When compared with the widely used dynamic comparators, the continuous type is a much more energy demanding component, although it provides the capability to be always tracking the input changes \cite{Sansen2006}.

The work in \cite{Marisa2017} tries to use dynamic comparator as means to keep track of the input signal, but it turns the level crossing architecture into a clocked system which gives away on the purpose of being efficient in the conversion of sparse signals. Nonetheless, the improvement in power efficiency in that work using the nominal voltage of the technology is noteworthy.

\begin{figure}[!t]
	\centering
	\includegraphics[width=1.0\linewidth]{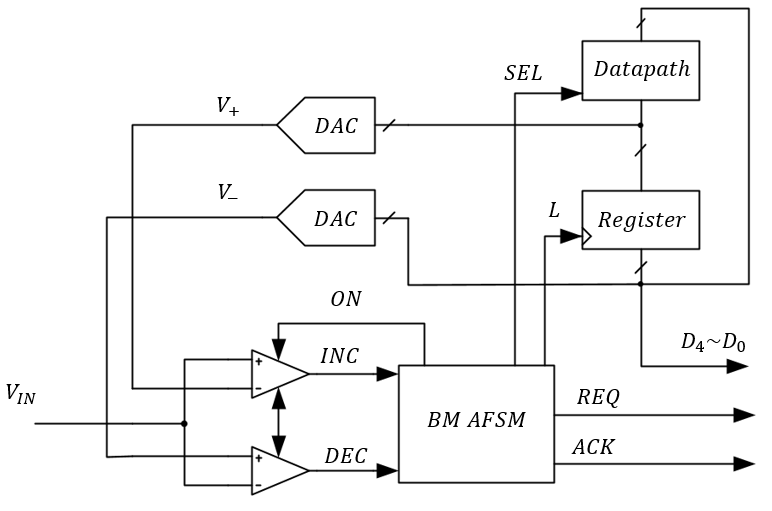}
	\caption{Proposed floating window architecture with added signal to turn off the comparators.}
	\label{adc_architecture}
\end{figure}

This work goes further in the investigation of adapting the comparators as a means to reduce the average power consumption. For that matter, it was used continuous comparators that could be turned off at given time so the static power is minimized. To enable this control over the comparator, the modified floating window architecture in Figure \ref{adc_architecture} was implemented. 

\begin{figure}[!t]
	\centering
	\includegraphics[width=1.0\linewidth]{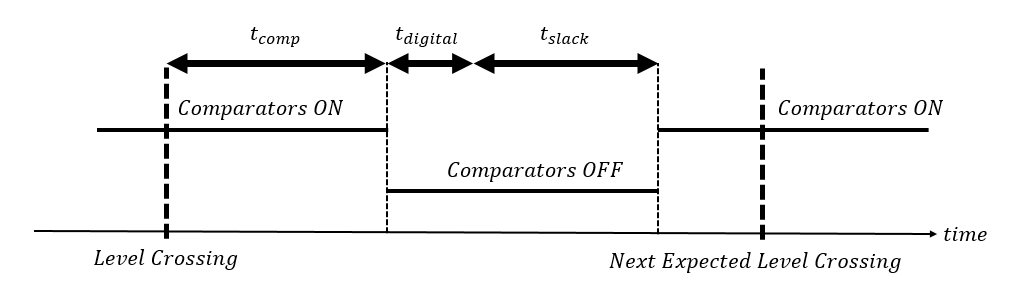}
	\caption{Timing of the events within two expected level crossing in the ADC.}
	\label{crossing_time}
\end{figure}

One can understand a level crossing event in this architecture as follow. At first, the signal $ V_{IN} $ is within the window formed by $ V_{+} $ and $ V_{-} $, then suppose an upper level crossing occurred, that is, the signal $ V_{IN} $ became bigger than $ V_{+} $. This will trigger the signal INC which feeds the Burst Mode Asynchronous Finite State Machine (BM AFSM). This state machine immediately turns off both comparators and proceeds to update the datapath and register holding the ADC conversion level at the moment. The changes in the register will update the value of the DACs, in this case for an upper output value, and the floating window moves up as it was supposed to be. The timing for the update of the ADC is set by the communication signals REQ and ACK on the interface with the state machine. During this whole update phase, the comparators will be off, so maximizing the time they are off will provide energy savings. However, the comparators cannot be definitely off since they need to be active before the next expected level crossing, thus the boundaries for the off time need to be set and the energy saving will be limited by it. Figure \ref{crossing_time} show the time representation of the comparator states and the events within the level crossing ADC. Note the time to update the digital components is much smaller than the time of decision from the comparator (timing in the figure is not scaled).

\section{ADC design}

\subsection{Comparator}
The comparators implemented in this work are based on the works of \cite{Zhang2014} and \cite{Bazes1991}. The self-bias comparator proved its low power performance in \cite{Zhang2014}. This component is composed of three stages, the first is a rail to rail differential input pair with folded cascode load. The second stage is a differential pair with a cross-coupled load to increase the overall gain of the comparator. Finally, the third stage is a modified positive feedback latch used to add the last bits of gain and to convert the differential signal to a single ended. Analytic gain equations could be taken for the stages, but with a moving bias point, it is hard to accurately calculate the gain of the comparator without a simulation tool. However, general amplifiers guidelines (\cite{Sansen2006}) are used to size the transistors and the final result is within the expected performance.

\begin{figure*}[!t]
	\centering
	\subfloat[First stage]{\includegraphics[width=0.32\linewidth]{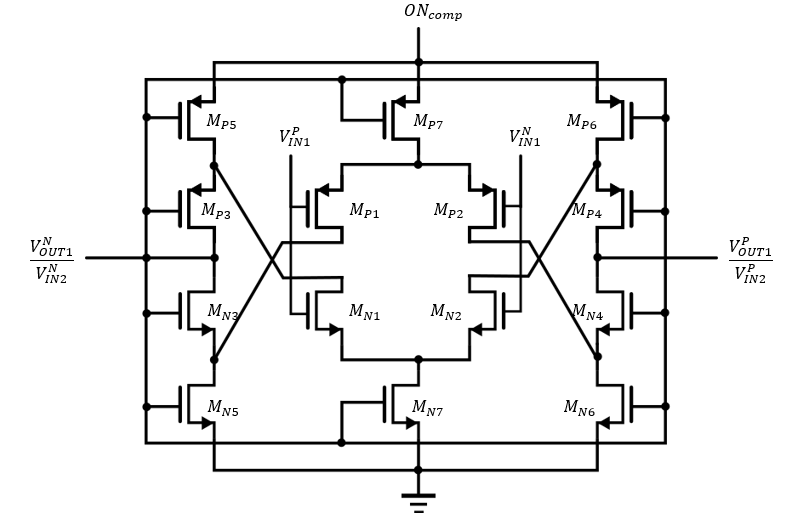}}
	\hfil
	\subfloat[Second stage]{\includegraphics[width=0.32\linewidth]{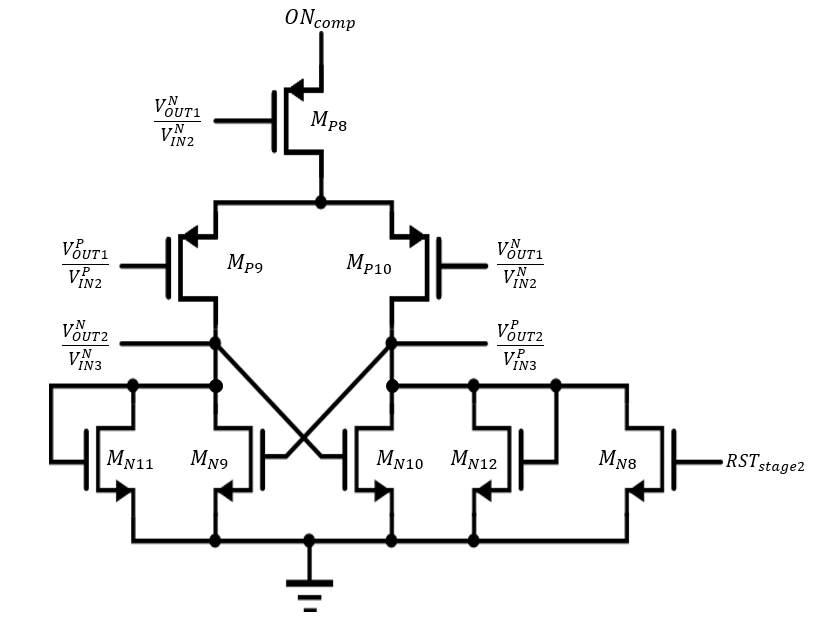}}
	\hfil
	\subfloat[Third stage]{\includegraphics[width=0.32\linewidth]{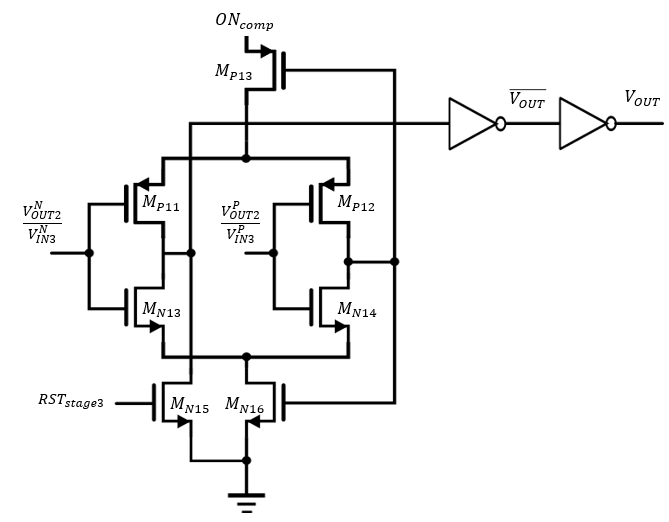}}
	\caption{Three stages of the self biased comparator used to track the input signal.}
	\label{comparator_stages}
\end{figure*}

The main difference is with respect to the measures taken to provide the turning off capability. Figure \ref{comparator_stages} provides the schematic for the three stages of the comparator. Note the main difference with the original comparator are the transistors $ M_{N8} $ and $ M_{N15} $, which both serve as discharging transistors to effectively prepare the comparator to be turned on again without causing glitches at the output. The signal $ ON_{comp} $ is a buffered version of the ON signal from the BM AFSM, in a way it can provide enough current to bias the comparator, and it also has the function of a voltage supply for the stages of the comparator.

Post-layout simulations show the typical decision time for the comparator to be around 659.5ns. With such value in hands, it is possible to calculate the expected bandwidth of the level crossing ADC using the Bernstein Theorem and the Equation \ref{eq_band} derived from \cite{Allier2003}. Where $ \Delta $ is equal to one quantization level and A is equal to half of the full scale voltage. This gives rise to a bandwidth of 11kHz for an ADC working at the full scale of 1V with such comparator.

\begin{equation}
\label{eq_band}
f_{max} = \frac{\Delta}{A \times 2\pi \times t_{comp}} 
\end{equation} 

Note this is not the equivalent of a Nyquist frequency since this is not a uniform sampled environment. In fact, the bandwidth here will be the maximum frequency up to which the ADC can still reliably track the input signal at full scale.

\subsection{Digital components}

\begin{figure}[!t]
	\centering
	\includegraphics[width=0.8\linewidth]{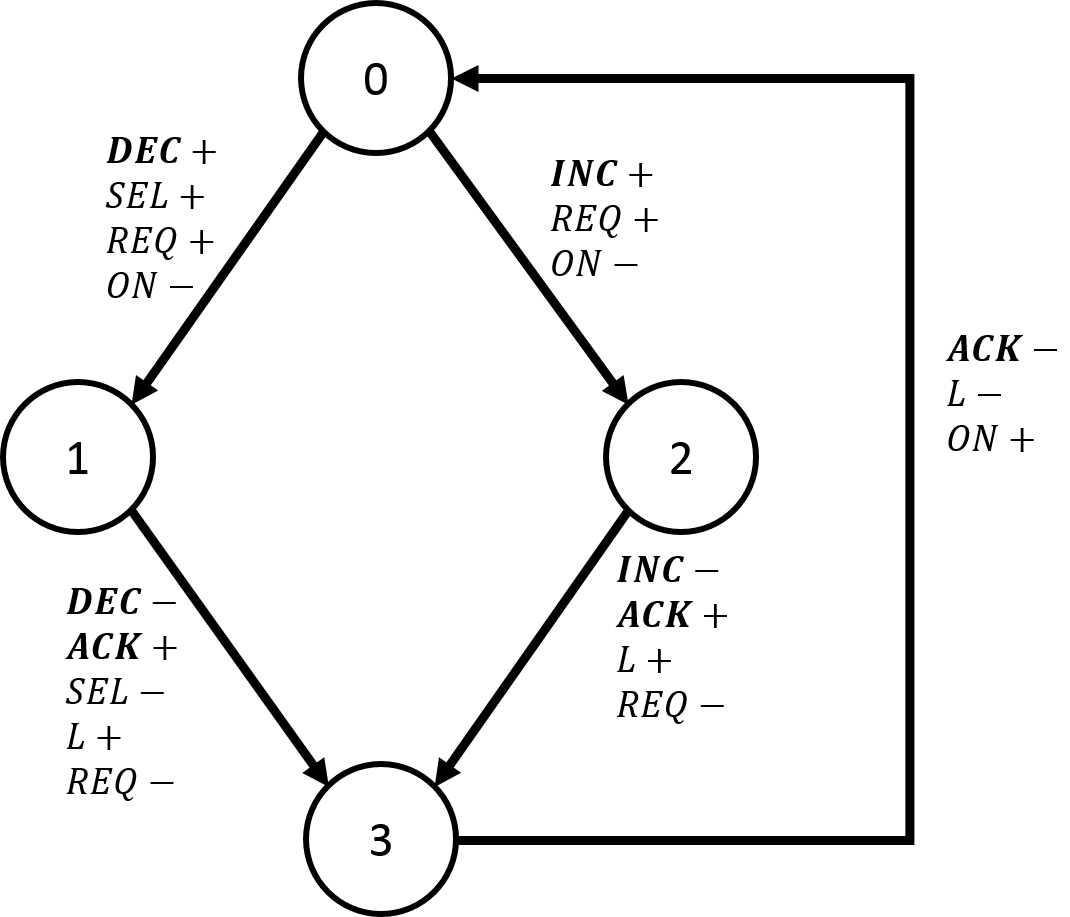}
	\caption{Graph of the state transitions of the BM AFSM.}
	\label{afsm_graph}
\end{figure}

The main concern in the digital part of the ADC is the state machine. This machine was chosen to be designed using the burst mode asynchronous paradigm \cite{minimalist1999}. This paradigm is meant for monotonic signal transitions in event-based environments, such as the one presented in the level crossing scheme of this work. The graph of the state transitions describing the behavior of the signals is shown in Figure \ref{afsm_graph}. The ``$ + $'' sign refers to a signal transitions from logic 0 to logic 1 and a ``$ - $'' sign refers to a signal transitions from logic 1 to logic 0. The signals INC, DEC, and ACK are inputs for the state machine, while the signals ON, REQ, SEL and L are the output. State 0 is the tracking state where the comparators are active. The path $ 0 \rightarrow 1 \rightarrow 3 \rightarrow 0 $ is triggered by a falling level crossing and the path $ 0 \rightarrow 2 \rightarrow 3 \rightarrow 0 $ is triggered by a rising level crossing. Note the time the ADC spends on the states 1, 2 and 3 contribute to the power savings since the comparator are off while in those states.

The state machine was synthesized using the 3D tool \cite{Yun1992} in the feedback output architecture, where the outputs of the state machine are used as state variables, thus saving area and possibly power. The resulting Boolean equations are shown in the set of Equations \ref{afsm_equations}.

\begin{equation}
\label{afsm_equations}
\begin{aligned}
REQ &= INC + DEC + \overline{ACK}.REQ \\
ON &= \overline{INC + DEC + ACK + REQ} \\
SEL &= DEC + \overline{ACK}.SEL \\
L &= \overline{INC}.\overline{DEC}.ACK
\end{aligned}
\end{equation}

To interface with the behavior of the BM AFSM, the designed register is a regular DFF based register working in the positive edge of the load signal L. The datapath is a custom ripple carry adder where the SEL signal swaps the second operand to provide an increment by 1 or to provide a decrease by 1 depending on what is needed. Note the standard value of SEL during track mode gives the datapath an output of $ +1 $ over the conversion level output (register value) and that is used to feed the DAC generating the $ V_{+} $.

\subsection{DAC}

\begin{figure}[!t]
	\centering
	\includegraphics[width=0.8\linewidth]{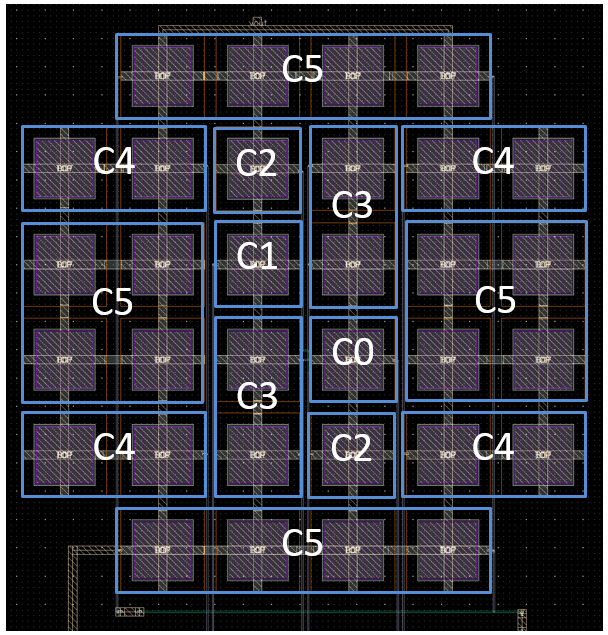}
	\caption{Layout of the capacitor banks of the CDAC.}
	\label{dac_layout}
\end{figure}

The DACs generating the voltage references were chosen to be CDACs based on charge sharing. This type of DAC has good performance and matching in medium resolutions \cite{VandePlassche2003} and has virtually no static power consumption. The focus of the design is given to the layout of such component. Based on the work of \cite{Jaworski2015} the floor planning was designed as shown in Figure \ref{dac_layout} aiming to minimize the routing parasitic and to increase the capacitors' matching.

\subsection{ACK generator}

\begin{figure}[!t]
	\centering
	\includegraphics[width=1.0\linewidth]{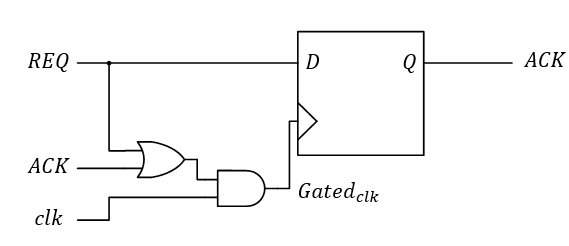}
	\caption{Controlled ACK generator for the communication with the BM AFSM.}
	\label{gated_clk_schematic}
\end{figure}

\begin{figure}[!t]
	\centering
	\includegraphics[width=1.0\linewidth]{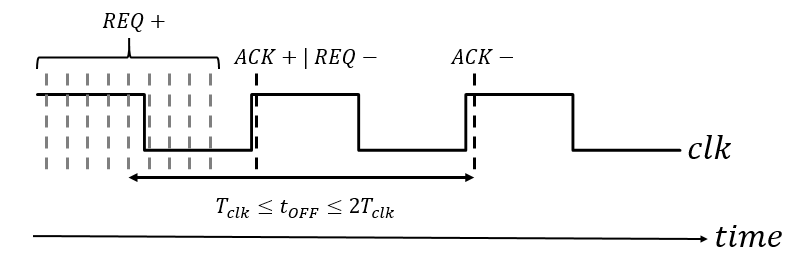}
	\caption{Timing for the behavior of the ACK generator after a REQ signal.}
	\label{gated_clk_timing}
\end{figure}

One of the simplest ways to provide the ACK signal to the BM AFSM in a controlled manner is shown in Figure \ref{gated_clk_schematic}. The clock-gating scheme is meant to decrease the power consumption during the self-transitions since the global clock signal does not arrive at the flip-flop in that condition \cite{Wu2001}. In Figure \ref{gated_clk_timing}, one can see the timing of the signal within the ACK generator and further extract the energy savings that it adds to the system. First, we consider a REQ+ signal just occurred, that means the comparators are off and the BM AFSM is waiting for an ACK+ signal. This waited response happens in the next rising edge of the input clock of the generator, taking the BM AFSM to a state where it keeps waiting for the ACK-- signal to go back to tracking mode. This signal is given at the second rising edge after the REQ+ occurred. Note that, even though a clock signal is being used for the ACK generator as means to control the time the ADC will be with the comparators off, the converter itself did not turn into a synchronous one, but rather it is using a clocked communication.

The energy saving estimate can be extracted from the fact that, given a big number of level crossing, the distribution of the moment when REQ+ occurs within a clock period becomes uniformly distributed. So the expected value from the moment a REQ+ occurs to the moment an ACK-- is received back is equal to the expected value of the uniform distribution plus one clock period. This time is the estimate of the time the comparators are off. With this value in hand, the set of Equations \ref{eq_gated_clock} models the average power obtained by measuring the power of the ADC for a long time ($t$).

\begin{equation}
\label{eq_gated_clock}
\begin{aligned}
&\hat{t}_{OFF} = \frac{3 \times T_{clk}}{2} \\
&T_{OFF} = N_{crossings} \times \hat{t}_{OFF} \\
&P_{ADC}^{mean} = P_{ADC}^{ON} \times (1 - \frac{T_{OFF}}{t}) + P_{ADC}^{OFF} \times (\frac{T_{OFF}}{t}) \\
\end{aligned}
\end{equation}

The increase in the period of the clock leads to more time in an off state and smaller average power. However, the value of this clock period is bounded by the frequency content of the input signal following the Equation \ref{eq_gated_bound}. This bound comes from the fact that the ACK-- signal has to arrive before the next expected level crossing, thus the worst case level crossing has to be within two clock periods to accommodate the communication time.

\begin{equation}
\label{eq_gated_bound}
T_{clk} \leq \frac{\Delta}{4\pi \times f_{IN} \times A} 
\end{equation}

\begin{figure}[!t]
	\centering
	\includegraphics[width=1.0\linewidth]{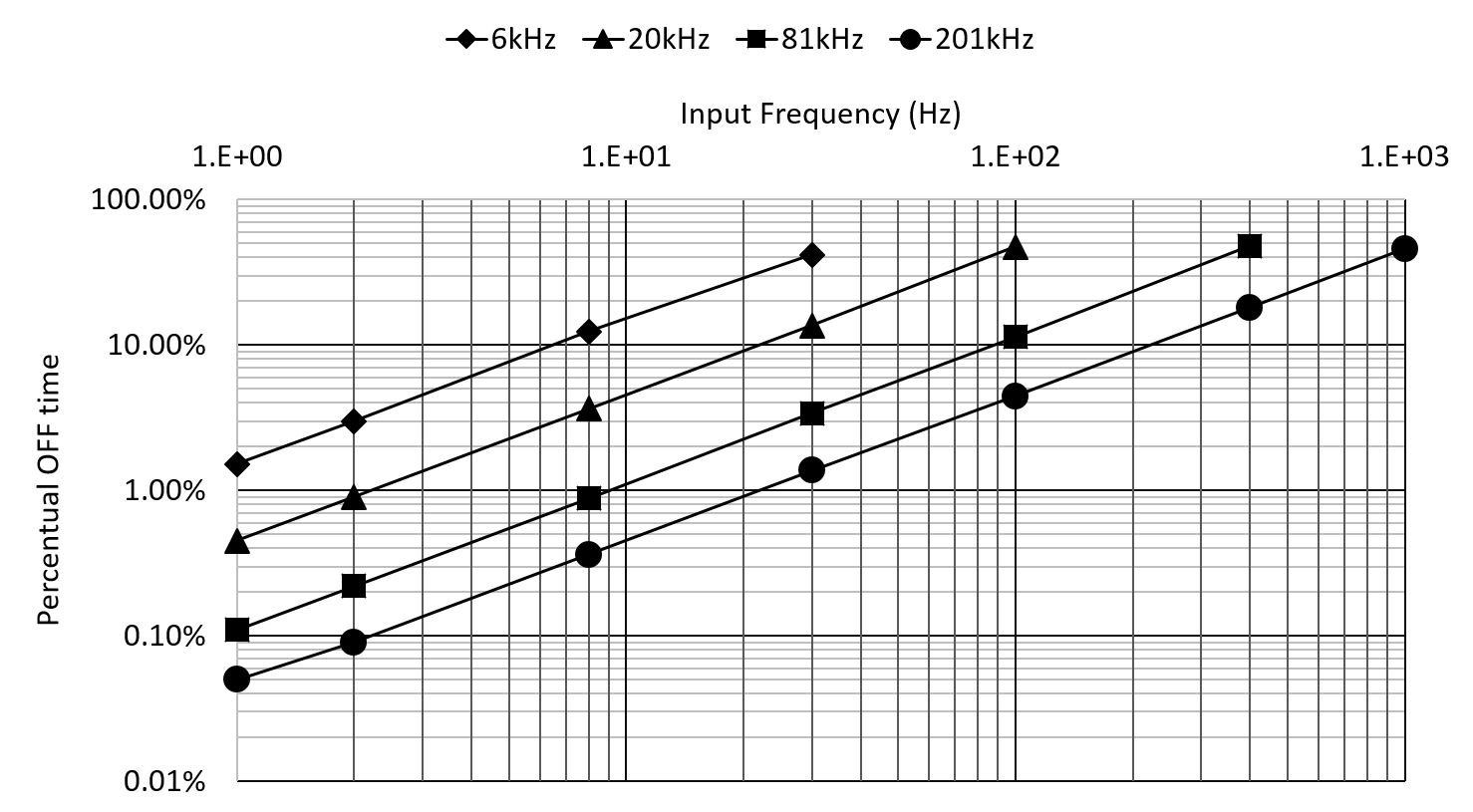}
	\caption{Plot of theoretical off time for different clock frequencies (curves) and different input frequencies (x-axis).}
	\label{octave_sinewave}
\end{figure}

Modeling this behavior in a numerical calculation software such as Octave, it is possible to plot the practical limits for power reduction with a set of sine waves as the input. Figure \ref{octave_sinewave} shows that for frequencies up to 1kHz, the amount of time the ADC will be off during one period using the maximum allowed clock period is close to 46.2\%.

For testing purposes, this component to generate the ACK signal for the ADC was not embedded within the integrated circuit, but rather it was modeled outside of the chip.

\section{Measurements and Analysis}

\begin{figure}[!t]
	\centering
	\includegraphics[width=1.0\linewidth]{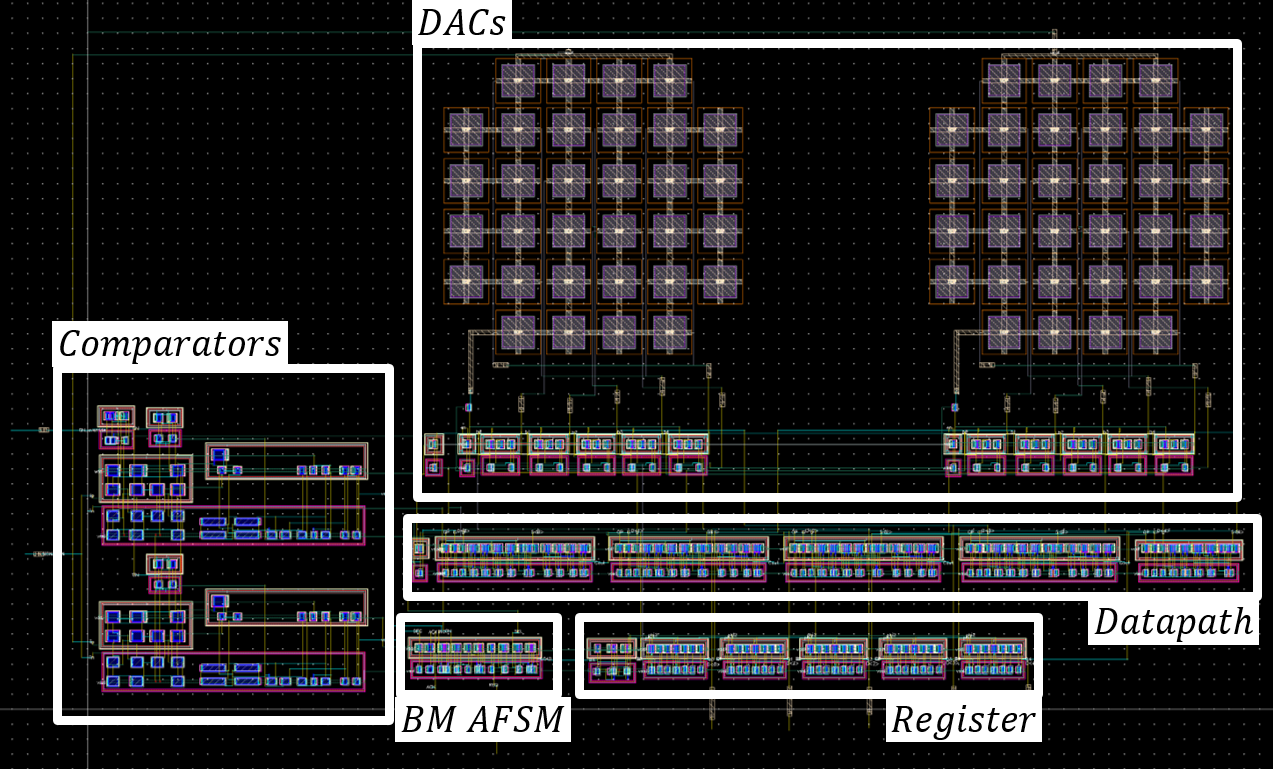}
	\caption{Layout of the ADC occupying an area of $0.05mm^2$.}
	\label{adc_layout}
\end{figure}

\begin{figure}[!t]
	\centering
	\includegraphics[width=1.0\linewidth]{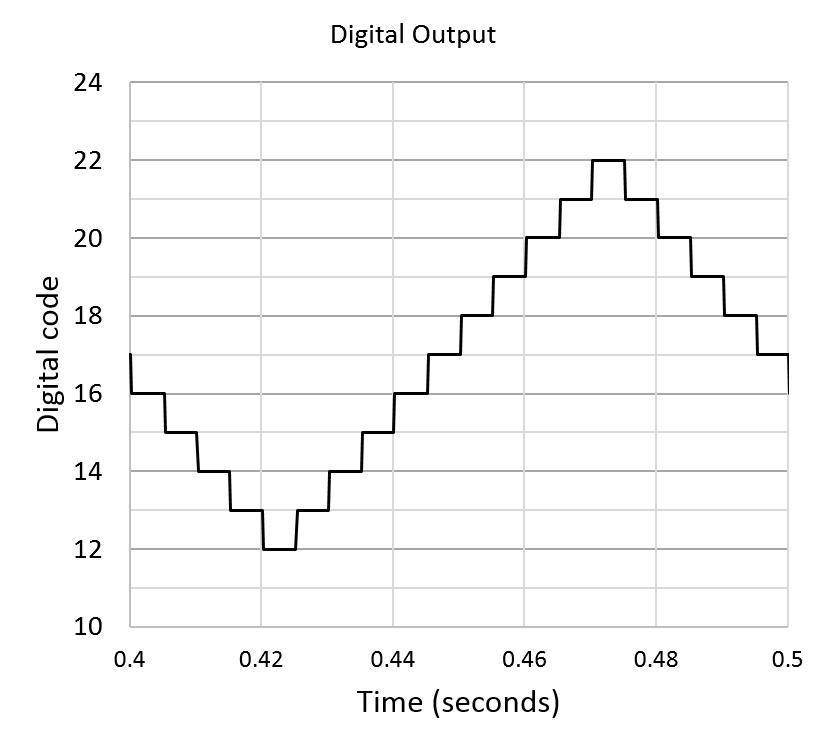}
	\caption{Plot of the output of the measured ADC during one clock period.}
	\label{adc_out}
\end{figure}

\begin{figure}[!t]
	\centering
	\includegraphics[width=1.0\linewidth]{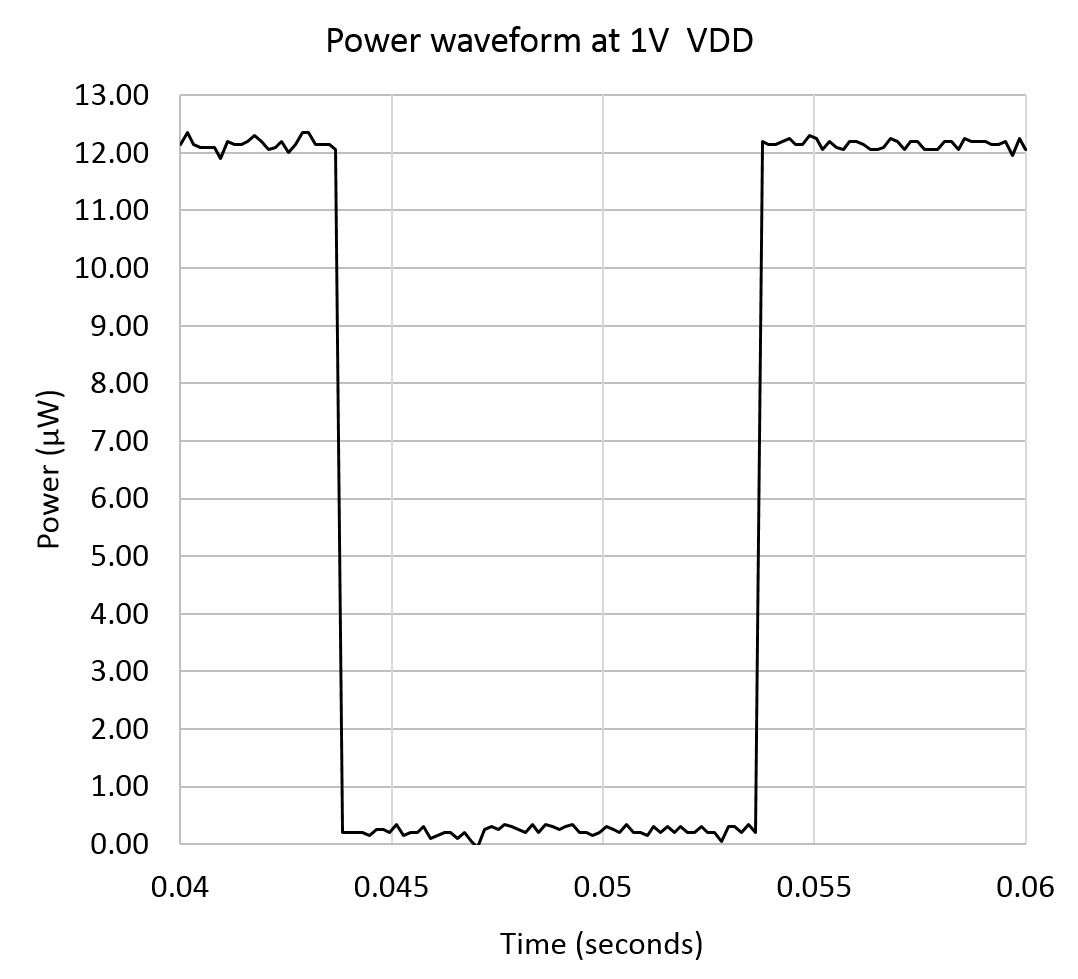}
	\caption{Magnified plot of the measured ADC power consumption.}
	\label{adc_power}
\end{figure}

The layout for the integrated circuit is shown in Figure \ref{adc_layout}. It was fabricated using the TSMC $0.18\mu m$ technology and it occupies an area of $0.05mm^2$. For measurement purposes, the ADC was subject to a slow triangular wave with limited amplitude and DC offset close to the middle voltage of conversion ($V_{DD}/2$). This was meant to ease the communication with the ADC and enable the precise collection of the on and off state power. Figure \ref{adc_out} shows the output waveform of the ADC under these conditions while Figure \ref{adc_power} shows a magnified version of the power waveform of the measured ADC.

With the extracted performance of the ADC, it was possible to evaluate the performance of the power reduction technique under the conversion of physiological signal using the ECG database in \cite{Goldberger2000} and the toolbox in \cite{Silva2014}.

\begin{figure}[!t]
	\centering
	\includegraphics[width=1.0\linewidth]{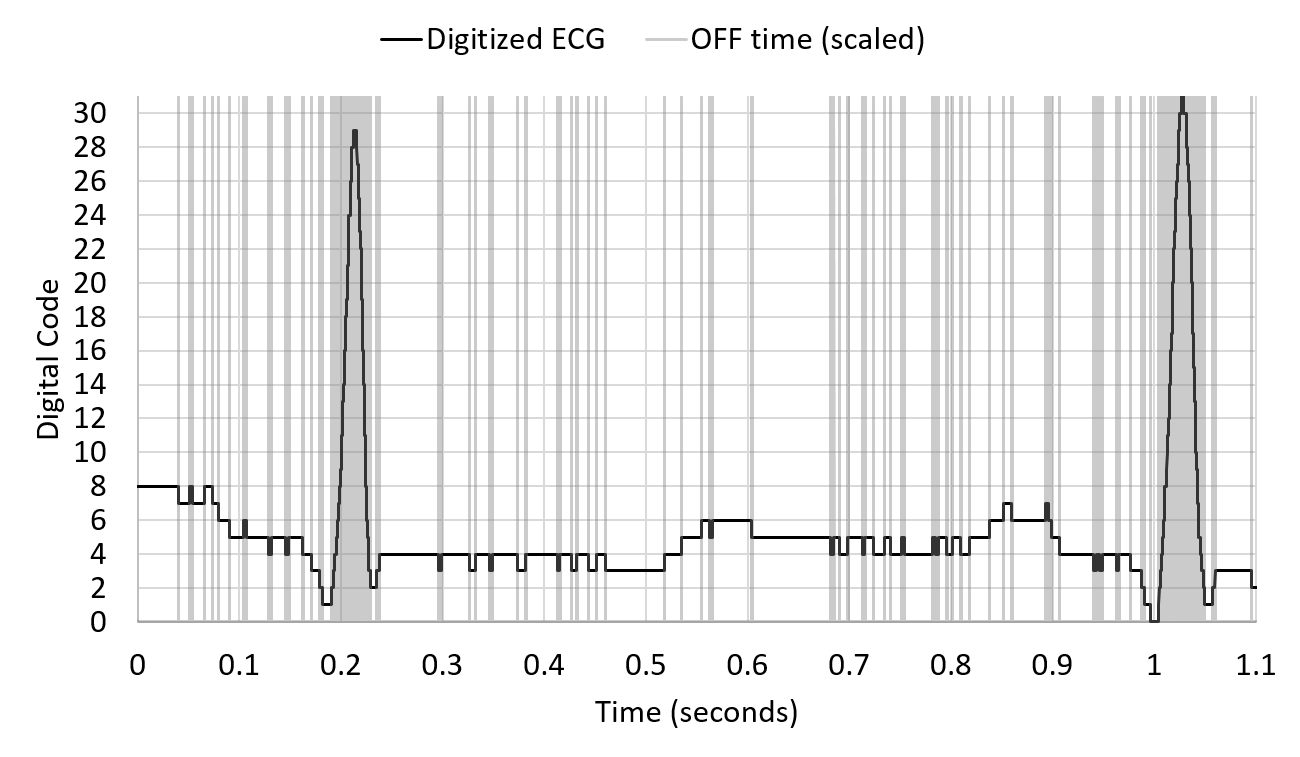}
	\caption{Magnified plot of the estimated off time of the ADC overlapped with the conversion of the ECG signal.}
	\label{ecg_octave}
\end{figure}

Under an ECG signal boundary, the previous timing analysis was made to design the clock period used to drive the system that generates the ACK signal to the ADC. Then, the ADC model was subject to a full scale ECG and the off time was measured with the modelling tool. A frame of such evaluation is shown in Figure \ref{ecg_octave} with the darker signal showing the moments the ADC was turned off. The average power reduction of the ADC was close to 4.2\% for the whole frame, and the peak reduction was close to 32.2\% when the converter is subject to the most activity, that is, when it is converting the QRS spike of the ECG signal.

Although the level crossing ADC is meant for a low power behavior during idle time in sparse signals, these results shows how the power consumption can be greatly reduced during high activity moments of conversions, increasing the usability of the level crossing ADC.

Table \ref{tb_comparison} shows the performance of different level crossing ADCs from the last years. The bigger average power consumption of the fabricated ADC is due to its increased bandwidth, which depending on the application could be reduced to further save energy. Also, the reduction in power in \cite{Zhang2014} and \cite{Li2013} is boosted by the reduction of the nominal voltage, what does come with the penalty of reduced dynamic range. 

\begin{table*}[!t]
\renewcommand{\arraystretch}{1.3}
\caption{Performance comparison of recent level crossing ADCs}
\label{tb_comparison}
\centering
\begin{tabular}{|c|c|c|c|c|c|c|}
\hline
		Reference				& Bits & Window & Technology & $ V_{DD} $ & Power & Bandwidth\\
		\hline
		\cite{Grimaldi2011}		& 10 & Floating & $ 0.15\mu m $ & $ 1.8V $ & $ 1300\mu W $ & $ 5kHz $\\ 
		\cite{Li2013}			& 6  & Fixed    & $ 0.18\mu m $ & $ 0.8V $ & $ 0.5\mu W $  & $ 5kHz $ \\ 
		\cite{Zhang2014}		& 5  & Floating & $ 0.13\mu m $ & $ 0.3V $ & $ 0.2\mu W $  & $ 1kHz $ \\ 
		\cite{Marisa2017}		& 4-8& Floating*& $ 0.35\mu m $ & $ 1.8V $ & $ 2\mu W $    & $ 1kHz $ \\
		This work 				& 5  & Floating & $ 0.18\mu m $ & $ 1.0V $ & $ 6.7\mu W - 12.2\mu W$ & $ 1kHz $ \\
		\hline
		\hline
		\multicolumn{7}{l}{*synchronous level crossing}  \\
		\hline 
\end{tabular}
\end{table*}

\section{Conclusion}

The proposed technique to reduce the power of an asynchronous level crossing ADC by means of turning off the comparators during idle time was implemented and showed potential in reducing the average power by maximizing the off time of the ADC. Simulations showed a limit of 45.5\% of power reduction for the conversion of pure sine waves and values smaller than that for ECG scaled waves, being the actual reduction a function of the activity of the signal. It was noted that the system still has margin to power savings by redesigning the comparator for smaller bandwidths and reducing the supply voltage in environments that allows reduced dynamic range. As for the next steps on the research, new topologies of comparators should be investigated to enable an easier control of the on/off time of the ADC with smaller static power consumption, such that the proposed technique could be ported to either floating window and fixed window level crossing ADCs with the lesser effort.

\section*{Acknowledgment}
The authors would like to thank professor Jacobus Swart and IMEC for making the fabrication of the IC possible through the program Mini-ASIC. This study was financed in part by the Coordena\c{c}\~ao de Aperfei\c{c}oamento de Pessoal de N\'ivel Superior - Brasil (CAPES) - Finance Code 001"

\ifCLASSOPTIONcaptionsoff
  \newpage
\fi



%

\bibliographystyle{./IEEEtran}
\bibliography{./IEEEabrv,./IEEEexample}

%
\vfill

\begin{IEEEbiographynophoto}{Lucas Moura Santana}
was born in Reden\c{c}\~ao, Brazil. He received his B.Sc degree in electronic engineering from the Technological Institute of Aeronautics - ITA, in 2017. Currently he is pursuing his M.Sc. degree in the area of low power analog-to-digital converters focused on bio-medical applications. His research interests are ADC, low power circuits, processing of physiological signals and synthesis of asynchronous circuits.
\end{IEEEbiographynophoto}

\begin{IEEEbiographynophoto}{Duarte de Oliveira Lopes} was born in Vila Real, Portugal. He received the B.Sc. degree in electrical engineering from the University of Mogi das Cruzes, Brazil, in 1980, the M.Sc. degree in computer science from the Technological Institute of Aeronautics, SJC, S\~ao Paulo, Brazil, in 1988, and the Ph.D. degree in electrical engineering from the Polytechnic School of the University of S\~ao Paulo in 2004. Since 2013, he is an Associate Professor in the Electronic Engineering Division at the Technological Institute of Aeronautics - ITA, Brazil. His research interests include asynchronous circuits, computer-aided digital design, low power digital systems, and logic synthesis. He is the author of more than 100 technical papers.
\end{IEEEbiographynophoto}


\begin{IEEEbiographynophoto}{Lester de Abreu Faria}
was born in Rio de Janeiro - RJ, Brazil. He has received degrees in aeronautical sciences from the Brazilian Air Force Academy, in 1993, and in electronic engineering from the Technological
Institute of Aeronautics-ITA, in 2004. He received the M.S. degree in electronic engineering
and computer science, in 2010, with emphasis on microelectronics, testability and design of analog read-out integrated circuits (ROICs). He received the Dr.Sci. degree in cryogenic integrated circuits for infrared detectors applications in 2014. He is currently a Pilot Colonel in the Brazilian Air Force and an Associate Professor of the Electronic Engineering Division, in Technological Institute of Aeronautics. His research interests include read-out integrated circuits for infrared detectors, cryogenic models for conventional MOS transistors, rad-hard
components and synthesis of asynchronous circuits.
\end{IEEEbiographynophoto}




\end{document}